\documentclass[submission,copyright,creativecommons]{eptcs}
\providecommand{\event}{LTT26} 

\usepackage{iftex}

\ifpdf
  \usepackage{underscore}         
  \usepackage[T1]{fontenc}        
\else
  \usepackage{breakurl}           
\fi

\newcommand{\ELO}{$\exists$loise}
\newcommand{\ABE}{$\forall$belard}
\usepackage{amsmath, amssymb, amsthm}
\usepackage{mathrsfs}
\newtheorem{definition}{Definition}
\newtheorem{lemma}{Lemma}

\newtheorem{theorem}{Theorem}

\title{Two Remarks about Game Semantics of Classical Logic}
\author{Thierry Coquand
\institute{Computer Science and Engineering\\
University of Gothenburg\\
Gothenburg, Sweden}
\email{coquand@chalmers.se}
}
\def\titlerunning{Two Remarks about Game Semantics}
\def\authorrunning{Thierry Coquand}
\begin{document}
\maketitle

\begin{abstract}
We present and explain two unpublished remarks of Stefano Berardi connected to game semantics.
\end{abstract}

\section{Introduction}

Around 30 years ago, I had several discussions with Stefano Berardi on the topic of game semantics
for classical logic, in particular connected to our work \cite{berardiBC95}. Stefano had several
insightful remarks, unfortunately most of them unpublished, and the goal of this note is to report
and comment on two of these remarks:
\begin{enumerate}
\item The first remark suggests a natural extension of the notion of views and debates \cite{coquand92,coquand95}
  to {\em transfinite} interaction sequences
\item The second remark shows that the game motivation of realization semantics \cite{berardiBC95} also validates
  {\em false} formulae, a remark which connects to recent work \cite{soloviev24}
\end{enumerate}
We first recall the main notions for the game semantics
used in \cite{berardiBC95} and then present these remarks.

\section{Game semantics of classical logic}

The semantics is defined for an infinitary propositional calculus. The formulas of this calculus are defined inductively as: (i) 0 and 1 are (atomic) formulas, and (ii) if $a_{i}$ $(i\in I)$ are formulas, where $I$ is a countable set, then both $\wedge_{i}a_{i}$ and $\vee_{i}a_{i}$ are formulas. Note that each arithmetical formula can be represented as a formula of this infinitary propositional calculus in a natural way.
Here, we regard atomic formulas as both universal and existential.  We define $\neg a$ by induction on $a$, using de Morgan rules.

We can define intuitionistic validity, specifying the set $\mathcal{V}$ of intuitionistically valid formulas inductively: (i) $1\in \mathcal{V}$,
(ii) $\wedge_i a_i\in \mathcal{V}$ if $a_i\in\mathcal{V}$ for all $i$ and (iii) $\vee_i a_i\in \mathcal{V}$ if $a_i\in\mathcal{V}$ for some $i$.
We can consider the formula as specifying a perfect information game and then intuitionistic validity corresponds to a winning strategy for this game. Note that we have a winning strategy for $a\vee \neg a$ for any $a$, by induction on $a$, by a ``copy-cat'' strategy.

We then introduce the notion of classical validity by specifying the set $\mathcal{C}$ of classically valid formulas. $\mathcal{C}$ is defined inductively:
(i) $1 \in \mathcal{C}$, (ii) $\wedge_{i } a_i \in \mathcal{C}$ if $a_i \in \mathcal{C}$ for all $i $, and (iii) $\vee_{i } a_i \in \mathcal{C}$ if there exists an $i_0 $ such that either $a_{i_0}$ is 1, or $a_{i_0}$ is of the form $\wedge_{j} a_{i_0 j}$ with $a_{i_0 j} \lor \vee_{i } a_i \in \mathcal{C}$ for all $j \in J$.

Game theoretical semantics for this calculus is also given as a perfect information game over a formula between two players: \ELO,
who plays for existential formulas, and \ABE, who plays for universal formulas. But the difference is now that \ELO~ can backtrack.
The game for a formula $a$ is played as follows: If \ELO~ (resp. \ABE) has to play and $a$ is atomic, then
\ELO~ (resp. \ABE) wins if $a$ is 1 (resp. 0). If $a$ is universal of the form $\wedge_{i } a_i$, then \ABE~ has to
choose an $i \in I$ and \ELO~ starts the game for $a_i$. If $a$ is existential of the form $\vee_{i } a_i$, then \ELO~
chooses an $i \in I$ and wins if $a_i$ is 1, loses if $a_i$ is 0. When $a_i$ is universal of the form
$\wedge_{j} a_{ij}$, \ELO~ can start the game not for $a_{ij}$ but for $a_{ij} \lor \vee_{i } a_i$
after \ABE~ returns a $j \in J$. It is only \ELO~ who is allowed to change her mind
and backtracks in her choice. The intuition is that \ELO~ learns from the environment \ABE, by playing in this way.

For instance, if $f$ is a function given as an oracle, the formula $\vee_x \wedge_y f(x)\leqslant f(y)$, stating that $f$ takes a minimum value,
is {\em not} intuitionistically valid. However \ELO~ has a winning strategy for the classical game. She chooses first an arbitrary value $x = 0$.
If \ABE~ answers with a value $y = x_1$ such that $f(0)\leqslant f(x_1)$, then \ELO~ wins. Otherwise $f(x_1)<f(0)$ and \ELO~
backtracks by choosing $x = x_1$. 
If \ABE~ answers with a value $y = x_2$ such that $f(x_1)\leqslant f(x_2)$, then \ELO~ wins. Otherwise $f(x_2)<f(x_1)$ and \ELO~
backtracks by choosing $x = x_2$, and so on. The game has to finish eventually since we have $f(x_n)<f(x_{n-1})$.
Note that \ELO~ may win without having found {\em the} actual minimum for $f$. 

One main difference with Lorenzen's approach is that we limit ourselves to $\wedge \vee$ formulae\footnote{In this way, the intuitionistic strategies
have a direct perfect information game interpretation.}. The strategies with backtracking correspond then to cut-free proofs, and the contribution
w.r.t. Lorenzen's work is to analyse what corresponds to the process of {\em cut-elimination}. 

In general, we represent a formula as a $\wedge \vee$ tree, possibly infinitely branching, with leaves being $0$ or $1$. For instance:
\begin{center}
  $(\exists_n\forall_m\, f(n) \leqslant f(m)) \rightarrow \exists_u\, f(u) \leqslant f(u+1)$
\end{center}  
will be represented as a $\wedge \vee$ tree:
\begin{center}
  $(\wedge_ n\vee_m\, f(n) > f(m)) \vee \vee_u\, f(u) \leqslant f(u+1).$
\end{center}    
The strategy for \ELO~ is the following, winning with at most two moves
\begin{itemize}
    \item \ELO~ asks for $n$.
    \item \ABE~ answers $n = a$.
    \item If $f(a) \leqslant f(a + 1)$, then \ELO~ takes $u = a$.
    \item If $f(a) > f(a + 1)$, then \ELO~ takes $m = a + 1$.
\end{itemize}

The strategies considered so far correspond to \textit{cut-free proofs}, which describe how a proof behaves in an environment that does not change its mind.
The cut-rule is interpreted as "cooperation" between proofs. For example:
 the strategy $A$ for
 \begin{center}
   $\vee_n\wedge_m\, f(n) \leqslant f(m)$
 \end{center}
 interacts with the strategy $B$
 for
 \begin{center}
   $(\wedge_n\vee_m\, f(n) > f(m)) \vee \vee_u\, f(u) \leqslant f(u+1)$
 \end{center}
to produce a proof of $\vee_u\, f(u) \leqslant f(u+1).$ The cut-formula is $\vee_n\wedge_m\, f(n)\leqslant f(m)$.

For a simple example, consider the function: $f(0) = 10,~f(1) = 8,~f(2) = 3,~f(3) = 27,\ldots$

The interaction proceeds as follows\footnote{The map $\varphi$ indicates to what moves we answer, with $0$
as the start move.}
\begin{enumerate}
    \item $B$ asks for $n$, with $\varphi(1) = 0$
    \item $A$ answers $n = 0$, with $\varphi(2) = 1$
    \item $B$ responds with $m = 1$, since $f(0)>f(0+1)$, with $\varphi(3) = 2$
    \item $A$ backtracks and answers by playing $n = 1$, with $\varphi(4) = 1$
    \item $B$ responds with $m = 2$, since $f(1)>f(1+1)$, with $\varphi(5) = 4$
    \item $A$ backtracks and answers by playing $n = 2$, with $\varphi(6) = 1$
    \item $B$ concludes by playing $u = 2$, since $f(2)\leqslant f(2+1)$
\end{enumerate}

The cut-formula, viewed as a tree, serves as the ``topic of the debate.'' The debate consists of:
\begin{itemize}
    \item Arguments and counter-arguments.
    \item Two opponents who can \textit{both} change their minds.
    \item At any point, they can \textit{resume} the debate from a previous point.
\end{itemize}

In this example, $A$ and $B$ debate, learning both from this interaction, until $B$ can produce a value for $u$.
We can think of $A$ as acting as the proof while $B$ acts as a counter-proof. Note that, in this example, if e.g.
$f(4) = 0$, it happens that the interaction stops before $A$ finding the actual minimum of $f$.

Gentzen's cut-elimination corresponds to the fact that such a debate has to end eventually. In \cite{coquand92}, reproduced in
\cite{curien98} and  \cite{clairambaultH10} (which also gives an alternative proof of termination), we
gave an argument for termination, which we believe to be essentially different from the one of Gentzen\footnote{For instance, F. Aschieri, a former student
of Stefano Berardi, showed \cite{aschieri17} that one could use this analysis to refine Gentzen's bounds on cut-elimination, by taking
into account, not only the complexity of the cut-formula, but also the level of backtracking of the two strategies that are
debating. In \cite{coquand95}, we give a constructive version of termination.}.
This argument relies on a geometrical analysis of the interaction, in the form of an {\em interaction sequence}.

\begin{definition}
  An interaction sequence is given by a pair
$V, \varphi$ with $V(n)\subseteq [0,n[$ and $\varphi(n)\in V(n)$ for $n>0$ with the conditions $V(1) = \{0\}$ and
      $V(n+1) = \{n\}\cup V(\varphi(n))$ for $n>0$.
\end{definition}

Intuitively, $\varphi(n)$ records which earlier move the current move responds to.

For example, the interaction sequence (pointer structure) produced by this interaction between the strategy $A$ and $B$ above is
\[
\varphi(1) = 0, \quad \varphi(2) = 1, \quad \varphi(3) = 2, \quad \varphi(4) = 1, \quad \varphi(5) = 4, \quad \varphi(6) = 1.
\]

For a lively description of how such interaction sequence is obtained we refer to \cite{curien98} and \cite{aschieri17}.
Let us define the segment $S(k)$ to be $[\varphi(k),k]$ if $k>0$ and $S(0)$ to be $[0,0]$.
We note that, for each $n>0$, we have a partition of $[0,n[$ in segments $S(m_k)$ with $m_0 = n-1$
    and $m_{k+1} = \varphi(m_k) - 1$. We have $V(n) = \{m_0,m_1,\dots\}$, and this is the ``view'' at stage $n$,
    notion introduced in \cite{coquand92} which
has been later used in game semantics of programming languages\footnote{In \cite{coquand95}, we show that we can define operations that have
symmetry properties not simple to obtain with cut-eliminations; see \cite{neri2025} for a stochastic version of such a symmetric operation.}.

\section{First remark}

In \cite{coquand92}, we proved, using classical logic, that if we have an infinite interaction sequence
$V,\varphi$, then we can find an infinite sequence $n_k<n_{k+1}$ such that $\varphi(n_{k+1})= n_k +1$.
The proof uses the following observation, combined with an induction of the depth of the formula.
Define a segment $S(k)$ to be {\em definite} if $k$
is not in the image of $\varphi$.

\begin{lemma}\label{nest}
  The definite segments form a nest structure: if we have to distinct definite segments then either they are disjoint
  or one is well inside the other.
\end{lemma}

Stefano Berardi noticed that this result can be refined in the following way.

\begin{theorem} (classical)
  There is a {\em unique} sequence $n_k<n_{k+1}$ such that $n_k + 1 = \varphi(n_{k+1})$ and $S(n_k)$
  is a {\em partition} of $[0,\omega[$.
\end{theorem}

Uniqueness is essential since it indicates which of the two players can be considered as responsible for the infinite debate.

We don't give the proof, which is a simple variation of the argument in \cite{coquand92} relying on Lemma \ref{nest},
but instead expand on the significance
of this result. The infinite sequence $n_k$ should be seen as the {\em view} at stage $\omega$.
In this sequence, all the $n_k$ have the same parity. This means, intuitively, that if there is an infinite debate,
then we can blame exactly one of the two players. This player has then a view $V(\omega)$ given by the set $\{n_k\}$, and it should
then choose one $\varphi(\omega) = n_k$. We have then $V(\omega + 1) = \{\omega\}\cup V(n_k)$ and we can then  extend
{\em transfinitely} the interaction sequence. 

\section{Second remark}

In the work \cite{berardiBC95}, we gave a modified realizability interpretation of classical $HA^{\omega}$ (with some simplified
$A$-translation) extended with
countable choice. This was motivated by an extension of the previous game interpretation where we allow to play {\em functions}
and not only natural numbers.
For instance, countable choice will be represented by a formula
\begin{center}
  $\vee_f\wedge_x P(x,f(x))\vee \vee_x\wedge_y\neg P(x,y)$
\end{center}
The strategy for \ELO~ for countable choice is then the following
\begin{itemize}
\item \ELO~ plays an arbitrary function, for instance $f_0 = \lambda_n~0$
\item \ABE~ answers with a value $x = x_0$
\item \ELO~ backtracks and plays then $x = x_0$
\item \ABE~ answers with a value $y = y_0$
\item \ELO~ backtracks again and plays $f_1 = f_0,x_0\mapsto y_0$ (that is $f_0$ updated with the value $y_0$ for $x_0$)
\item \ABE~ answers with a value $x = x_1$
\item if $x_1 = x_0$, \ELO~ wins by playing $y= y_0$ and then playing
  $\neg P(x_0,y_0)$ against $P(x_0,y_0)$ by ``copy-cat'' strategy; otherwise \ELO~ backtracks and plays
  $x = x_1$, and so on
\end{itemize}
With this strategy, \ELO~ updates successively the values of $f$
\begin{center}
  $f_n = f_0,~x_0\mapsto y_0,~x_1\mapsto y_1,\dots, ~x_{n-1}\mapsto y_{n-1}$
\end{center}  
by asking \ABE~ what should be the value for $y$ as an answer to $x=x_i$.
If ever \ABE~ answers to $f = f_n$ by playing a value $x_n = x_i$ which has already been answered,
then \ELO~ wins by playing $y_n = y_i$ and then playing $\neg P(x_i,y_i)$ against $P(x_i,y_i)$.

In \cite{berardiBC95}, we remark that if \ABE~ answers in a ``continuous'' way, i.e. proceeds using only a finite amount of
information
about the function $f_n$ , it eventually has to answer to $f = f_n$ a value $x$ which has already been answered for some
$f_k,~k<n$. By the discussion above, this means that \ELO~ eventually wins in this case.

We used this strategy in \cite{berardiBC95}  to provide a modified realizability interpretation of classical $HA^{\omega}$ extended with
countable choice\footnote{Note that Spector \cite{spector:bar} also provided a computational interpretation of this system, but it was
using {\em Dialectica} interpretation and not modified realizability.}. In \cite{hida12}, T. Hida used a similar justification (by a continuity argument) for a strategy
for the {\em axiom of determinacy} and could also produce a modified realizability interpretation of classical $HA^{\omega}$ extended with
the axiom of determinacy\footnote{For more connected recent work, see \cite{provenzano2025}.}.

The second important remark of Stefano Berardi is the following: there are examples of {\em false} formulae of $HA^{\omega}$ which have
a strategy for \ELO~ winning against any {\em continuous} opponent. One such example is the following
\begin{center}
  $\vee_f\wedge_x\vee_y~f(x) =  0\wedge f(y) \neq 0~~~~~~~~~~~~~~~(*)$
\end{center}
Surprisingly, this formula, though false, admits a strategy $A_1$ for \ELO~ which is the following
\begin{itemize}
\item \ELO~ plays the function $f_0 = \lambda_n 1$
\item \ABE~ answers with $x = x_0$
\item \ELO~ backtracks by playing $f_1 = f_0, 0\mapsto 0$
\item \ABE~ answers with $x = x_1$
\item if $x_1 < 1$ \ELO~ wins by playing $y = 1$; otherwise \ELO~ backtracks and plays $f_2 = f_1,1\mapsto 0$
\item \ABE~ answers with $x = x_2$
\item if $x_2<2$ \ELO~ wins by playing $y = 2$; otherwise \ELO~ backtracks and plays $f_3 = f_2,2\mapsto 0$ and so on
\end{itemize}
If \ABE~ plays in a continuous way, it has to play $x_n < n$ at some point, and then \ELO~ wins by playing $y = n$.

But the formula $(*)$ is {\em false}, and we have a strategy $B_1$ for its negation
\begin{center}
  $\wedge_f\vee_x\wedge_y~f(x) \neq  0\vee f(y) = 0$
\end{center}
The strategy is as follows, after \ABE~ has played $f = g$
\begin{itemize}
\item \ELO~ plays an arbitrary value, e.g. $x = 0$
\item \ABE~ answers by playing $y = y_0$
\item if $g(0)\neq 0$ or $g(y_0) = 0$ then \ELO~ wins; otherwise $g(0) = 0$ and $g(y_0)\neq 0$
  and \ELO~ backtracks by playing $x = y_0$
\item \ABE~ answers by playing $y = y_1$, and \ELO~ wins since $g(y_0)\neq 0$
\end{itemize}

If we let $A_1$ play against $B_1$, we get an infinite debate, which corresponds to the fact that we
cannot expect to have cut-elimination.

\begin{enumerate}
    \item $A_1$ plays $f_0 = \lambda_n 1$, with $\varphi(1) = 0$
    \item $B_1$ answers $x = 0$, with $\varphi(2) = 1$
    \item $A_1$ plays $f_1 = f_0,0\mapsto 0$, with $\varphi(3) = 0$
    \item $B_1$ answers $x = 0$, with $\varphi(4) = 3$
    \item $A_1$ answers $y=1$, with $\varphi(5) = 4$
    \item $B_1$ plays $x=1$, with $\varphi(6) = 3$
    \item $A_1$ plays $f_2 = f_1,1\mapsto 0$, with $\varphi(7) = 0$
    \item $B_1$ plays $x = 0$, with $\varphi(8) = 7$
    \item $A_1$ answers $y = 2$, with $\varphi(9) = 8$
    \item $B_1$ answers $x=2$, with $\varphi(10) = 7$
    \item $A_1$ plays $f_3 = f_2,2\mapsto 0$, with $\varphi(11) = 0$, and so on.
\end{enumerate}

This debate will create the infinite sequence of functions
\begin{center}
  $f_0 = \lambda_n1,~f_1 = f_0,~0\mapsto 0,~f_2 = f_1,1\mapsto 0,~\dots$.
\end{center}

\medskip

It is thus remarkable that we could obtain a modified realizability
interpretation for countable choice; we cannot do it for $(*)$, despite the fact that $(*)$ also has a winning strategy for
\ELO~ against any continuous opponent. This shows that continuity-based arguments are insufficient by themselves to characterise
realizability semantics.

This example is to be compared with the work of S. Soloviev \cite{soloviev24}. If $f$ is the Ackermann function, and
\ABE~ is restricted to a primitive recursive strategy, then \ELO~ has a winning strategy for the (false)
formula
\begin{center}
  $\vee_x\wedge_y y\leqslant f(x)$.
\end{center}
The strategy consists in playing successively the values $x = 0,1,2,\dots$.
Since \ABE~ follows a primitive recursive strategy, it answers $y = y_0,~y_1,~y_2,\dots$ in a primitive recursive way
and we eventually should have $y_n\leqslant f(n)$. This example is philosophically similar to Stefano's second remark: if
there is a asymmetry between the two players (e.g. in computing resources) the stronger player can lead the other
player in believing a false statement.

\section*{Acknowledgement}

Many thanks to Stefano for so many fascinating discussions on type theory, constructive mathematics and computational content of classical proofs.
Thanks also to Sergei Soloviev for interesting more recent discussions on game semantics.

\bibliographystyle{eptcs}
\bibliography{coquandBiblio}

\end{document}